\documentclass [a4paper, 12pt]{article}
\usepackage{graphicx}
\usepackage[small]{subfigure,epsfig}
\usepackage {amsmath}
\usepackage{amssymb}
\usepackage{cite}
\usepackage{array,longtable,lscape}
\numberwithin{equation}{section} \numberwithin{table}{section}
\usepackage{lscape}

\begin{document}

\title{High order model for describing the pattern formation processes on semiconductor surfaces}
\author{Nikolay A. Kudryashov, Pavel N. Ryabov}

\date{\small{National Research Nuclear University MEPhI (Moscow Engineering Physics Institute), 31 Kashirskoe Shosse, 115409 Moscow, Russian Federation}}
\maketitle

\begin{abstract}
The self-organization processes of nanopattern formation on semiconductor surfaces under low energy ion beam bombardment is considered. The new mathematical model based on nonlinear  evolution equation of the sixth order is presented. The pseudo--spectral method for periodic boundary value problem for this equation is discussed. The influence of high order terms on the evolution of the semiconductor surface morphology is studied.
\end{abstract}



\section{INTRODUCTION}

Low energy ion bombardment of semiconductor substrates causes the self--organization processes of the nanoscale patterns on their surfaces \cite{Flamm_2001, Cuerno2001}. These artificially created patterns represent the ordered or unordered monodisperse structures in the shape of align ripples, quantum dots or holes formed depending on the experimental parameters. Nowadays this phenomenon is actively investigated since these structures are widely used in nanotechnology to upgrade and develop the new types of electronic devices with interesting and useful properties \cite{Cuerno2001, Gates2000}. Thus, together with experimental research, it is necessary to provide the theoretical study of the main features and regularities of the process of semiconductor surface erosion under low energy ion bombardment. One of the most interesting questions in that area is the derivation of mathematical models for describing the evolution of the surface roughness with time.

Currently, for prediction of the surface roughness of semiconductor surfaces under ion bombardment the continuum models are used. As an example of such model we can point out the linear Bradley--Harper model \cite{Bradley_1988} which was the first successful theoretical attempt allowing to study the formation of ripple patterns on the semiconductors under ion bombardment. However this model has a number of shortcomings. In particular it does not allow us to describe the mechanism of ripples stabilization and the formation of patterns in the form of nanodots or holes under a certain conditions \cite{Castro_2005}. Thus there we needed for a new model that would allow us to adequately describe the experimental results. As a the result, the Bradley--Harper model was replaced by the models of anisotropic Kuramoto--Sivashinsky type \cite{Makeev_2002, Cuerno2006, Cuerno2011}. Generally, most of these models have the fourth order since they take into account self-diffusion and thermal diffusion terms as most higher order terms in continuum equation. Thus, the question arises: does the higher order terms, like the fifth order dispersion and the sixth order diffusion, affect on the self-organization process of nanopattern formation? We are trying to answer on this question in the Letter.

The manuscript is organized as follows. In Section 2 we derive the new sixth order mathematical model for describing the nanopattern formation processes on the semiconductor surfaces under ion bombardment. In Section 3 we describe our numerical strategy and discuss obtained numerical results.

\section{DERIVATION OF CONTINUUM EQUATION FOR THE SURFACE HEIGHT}

\begin{figure}[!htb]
\center
\includegraphics[width=140 mm]{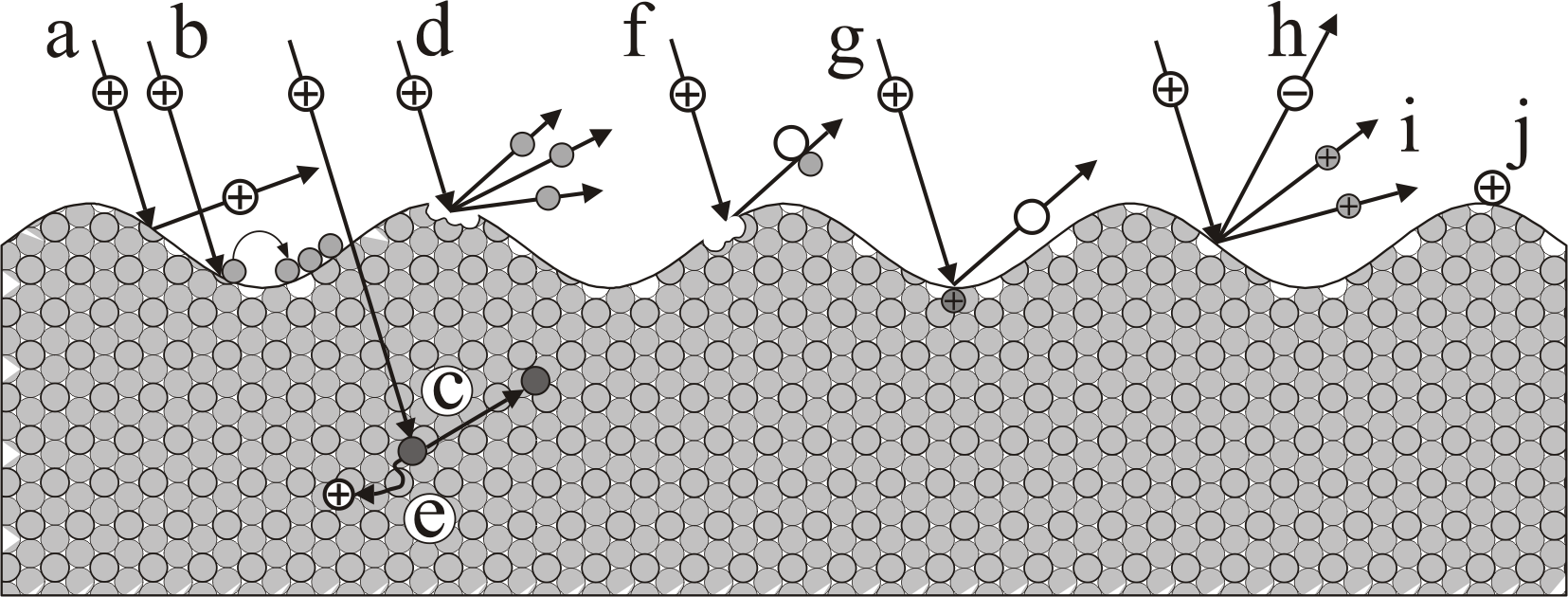}
\caption{{\fontshape{sl}\selectfont  Types of interaction between ions and the surface of substrates. (a) reflecting of an ion on atoms; (b) surface dislocations; (c) internal dislocations; (d) physical sputtering; (e) ion implantation; (f) chemical sputtering; (g) the charge transfer; (h) electron emission; (i) emission of surface ionized atoms; (j) charge transfer.}}\label{f}
\end{figure}

An interaction between low--energy ions with rigid bodies causes the sputtering processes of atoms from its surfaces and as a result, the surface morphology changes with time. Currently, these interactions can be developed in ten possible scenarios \cite{Brodie_1982}. According to the first scenario an ion can be reflected by an atom or a group of atoms, see Fig. \ref{f}(a). In the second scenario atomic dislocations on the surface may occur, i.e. transition of an atom from a state with a weak bond with crystallin lattice to a state with a strong bond as a result of collision with an ion that is illustrated in Fig. \ref{f}(b). Ions with enough energy can also cause the dislocation of atoms within the material (Fig. \ref{f}(c)) and knock atoms out of the body surface Fig. \ref{f}(d). This process is called the "physical sputtering". The ion, which gave all its energy and got deeper into the meter, can be captured by atoms from the crystallin lattice. This process is called the ion implantation and illustrated on Fig. \ref{f}(e). The atoms may also be in a gaseous state after the chemical reaction between ions and atoms. In this state, the binding energy of atoms with the surface is very low, so atoms can sputter from the surface, see Fig. \ref{f}(f). This sputtering mechanism is called the chemical sputtering. Typically, this type of interaction takes place when the energy of incident ions is $\epsilon \leqslant 0.5$ keV. Positive ions can capture electrons from the surface and then reflect from it in the neutral state (Fig. \ref{f}(g)). Ion bombardment of some metal surfaces can cause the secondary electron emission that is plotted in Fig. \ref{f}(h). This process may also be observed in the case of excitation of atoms to ionized state, see Fig. \ref{f}(i). Also it can be adsorbed by the surface (Fig. \ref{f}(j)).

In the present work we consider the processes of physical sputtering as a major scenario of interaction between the ions and the atoms of substrate.

The physical sputtering is related to the transfer of kinetic energy from the incident ions to the atoms of target. If the energy of incident ions exceeds the bond energy of atoms in crystallin lattice, the ions start to knock out atoms or small atom complexes from the surface. The overwhelming majority of the sputtering particles are neutral atoms. The proportion of ions does not exceed one percent.
If the energy of incident ions lies in the interval $10 \, \mbox{keV} \leqslant \epsilon\leqslant 100$ keV and their mass is more than the mass of the target atoms, the process of physical sputtering becomes dominant and most of the kinetic energy of ions transferred to the atoms concentrated in a superficial layer of a few nanometers thick.

The first theoretical explanation of the interaction mechanism between incident ions and target atoms was given  by Sigmund \cite{Sigmund_1969, Sigmund_1973, Sigmund_1981}. According to this theory ions penetrate into material and generate the cascades of atomic collisions in the surface layer. Since the most of the scattering atoms concentrate near the surface, the scattering occurs at some average depth $R$, which is called the average depth of ion penetration. The value of $R$ depends on the ion energy $\epsilon$
\begin{equation}\label{ch2:eq1}
R(\epsilon)=\frac{1-m}{2m}\gamma^{m-1}\frac{\epsilon^{2m}}{NC_m},
\end{equation}
where $N$ is a density of target atoms, $C_m$ is a constant, which depends on parameters of interatomic potential and $m(\epsilon)$ is a function varies in the interval $0\leqslant m\leqslant 1$. In our case, when $\epsilon\subseteq[10, 100]$, $m$ approximately equals to $1/2$ \cite{Sigmund_1969, Sigmund_1973, Makeev_2002}.

For the description of this rather complicated process of atomic collisions Sigmund suggested the following qualitative scenario of that process: the ions penetrate into material and stop at some internal point starting to emit their kinetic energy according to the Gaussian distribution
\begin{equation} \label{ch2:eq2}
E(\textbf{r})=\frac{\epsilon}{(2\pi)^{3/2}\sigma\mu^2}\exp\left\{-\frac{Z^2}{2\sigma^2}-\frac{X^2+Y^2}{2\mu^2}\right\},
\end{equation}
where $Z$ is a distance between the internal point to the surface point measured along the ion trajectory, $X, Y$ are perpendicular to it, $\epsilon$ is energy of the ions, $\sigma $ and $\mu$ are big and small semi-axis of the Gaussian energy distribution. Formula \eqref{ch2:eq2} is consistent  with results of experiments and numerical simulation, see for detail \cite{Eckstein_1991,Makeev_2002}.

Let $h(x,y,t)$ be a function that describes the evolution of substrate surface height. The equation relating the height function $h(x,y,t)$ with the erosion velocity $V_O$ at the surface point takes the form
\begin{equation}\label{ch2:eq3}
\frac{\partial h}{\partial t}=-V_O (1+h_x^2+h_y^2)^{1/2}-K\nabla^4h+\eta(x,y,t),
\end{equation}
where $K$ is a thermal softening parameter and $V_O$ is given by the relation
\begin{equation}\label{ch2:eq4}
V_O=\Lambda\int_{\Re}\Phi(\textbf{r})E(\textbf{r})d\textbf{r}.
\end{equation}
Where $\Lambda$ is a constant and depends on the scattering cross section and the surface binding energy, $\Phi(\textbf{r})$ is a correction of ions flux due to variation of local surface curvature and $\Re$ is a surface, each element of which makes a contribution to the total energy transferred by ions.

Taking into account \eqref{ch2:eq4} in \eqref{ch2:eq3} we obtain an integro-partial differential equation which describes the evolution of the surface height function with time. However this equation is hard to analyze. In the work by Makeev et.al. \cite{Makeev_2002}, the authors proposed an interesting method which allows us to reduce the integro-differential equation to the stochastic nonlinear partial differential equation of the fourth order with coefficients depending on the parameters characterizing the sputtering process. Then authors studied the morphological features that this model predicts and gave the experimental approvement of the results obtained.

As it was already mentioned, the forth order model was obtained in \cite{Makeev_2002}. In work \cite{Carter_1999} author noted that the high order terms(the fifth and more) change the pattern morphology. These terms change the wave number and growth rate by $10 \%$. However, the author mentioned that further studies are necessary. In the next series of works \cite{KRFK_2013, KudMig_2007, KudRyabSin_2011} it was shown that high order terms affect on the pattern formation processes in one dimension. In addition, we were motivated by \cite{Golovin, Nepomnyashchy}, where the process of crystal surface faceting by surface diffusion is studied. The mathematical model which was used by the authors has the sixth order. Thus using the technique proposed by Makeev et.al. \cite{Makeev_2002} we obtain the sixth order nonlinear partial differential equation for describing the pattern formation processes on semiconductor surface under ion bombardment in the following form
\begin{equation}\label{ch2:eq5}
\begin{gathered}
\frac{\partial h}{\partial t}=\mu_{1}\left(\frac{\partial h}{\partial x}\right)^2+\mu_{2}\left(\frac{\partial h}{\partial y}\right)^2+\nu_{1}\frac{\partial^2 h}{\partial x^2}+\nu_2\frac{\partial^2 h}{\partial y ^2}
+\xi_1\left(\frac{\partial h}{\partial x}\right)\left(\frac{\partial^2 h}{\partial x^2}\right)+\\+\xi_2\left(\frac{\partial h}{\partial x}\right)\left(\frac{\partial^2 h}{\partial y^2}\right)+\Omega_{1}\frac{\partial^3 h}{\partial x^3}+\Omega_{2}\frac{\partial^3 h}{\partial x \partial y^2}
-D_{1}\frac{\partial^4 h}{\partial x ^4}-D_{2}\frac{\partial^4 h}{\partial y ^4}-\\-D_{3}\frac{\partial^4 h}{\partial x^2 \partial y^2}+\zeta_1\frac{\partial^5 h}{\partial x^5}+\zeta_2\frac{\partial^5 h}{\partial x \partial y^4}+\zeta_3\frac{\partial^5 h}{\partial x ^3 \partial y^2}
-r_1\frac{\partial^6 h}{\partial x ^6}-r_2\frac{\partial^6 h}{\partial y ^6}-\\-r_{3}\frac{\partial^6 h}{\partial x ^4 \partial y^2}-r_{4} \frac{\partial^6 h}{\partial x ^2 \partial y^4}-K\nabla^4h+\eta(x,y,t),
\end{gathered}
\end{equation}
where parameters $\mu_{1,2}, \nu_{1,2}, \xi_{1,2}, \Omega_{1,2}, D_{1-3}$ are given in \cite{Makeev_2002, Bradley2011} and the parameters $r_{1-4}, \zeta_{1-3}$ are expressed in terms of physical quantities as
\begin{equation}\label{AppA:1}
\begin{gathered}
\zeta_{1}=\frac {R^{4}Fs}{120f^4}\left(c^2f\left(15(R_{\mu}^2-R_{\sigma}^2)+\frac{R_{\sigma}^{12} s^4 }{f^2}+\frac{R_{\sigma}^{12} s^{6}(R_{\mu}^2-R_{\sigma}^2)}{f^3}+\right. \right. \\ \left. \left.+\frac{10R_{\sigma}^8s^2}{f}+15R_{\sigma}^4+15\frac{R_{\sigma}^8 s^4(R_{\mu}^2-R_{\sigma}^2)}{f^2}+45R_{\sigma}^4 s^{2}( R_{\mu}^2-R_{\sigma}^2)\right)-\right.\\\left.-5(3f^2+R_{\sigma}^8s^4+6R_{\sigma}^4s^2f)\right),\\
\end{gathered}
\end{equation}
\begin{equation}\label{AppA:2}
\begin{gathered}
\zeta_{2}=-\frac{R^4FsR_{\sigma}^2}{8R_{\mu}^4f}\left(1-\frac{R_{\sigma}^2R_{\mu}^2c^2}{f}\right), \\
\end{gathered}
\end{equation}
\begin{equation}\label{AppA:3}
\begin{gathered}
\zeta_{3} = -\frac{R^{4}Fs}{12f^4R_{\mu}^2}\left( 3f^3\left\{1+\frac {R_{\sigma}^4 s^2}{f}\right\}
-R_{\sigma}^4 c^2f^2\left\{ 3+\frac {R_{\sigma}^4 s^2}{f} \right\}-\right. \\ \left. -c( R_{\mu}^2-R_{\sigma}^2) ( 3f^2+6R_{\sigma}^4 s^2 f+R_{\sigma}^8s^4)\right),\\
\end{gathered}
\end{equation}
\begin{equation}\label{AppA:4}
\begin{gathered}
r_{1}= \frac {FR^5R_{\sigma}^2}{720f^3}\left( \left\{15+\frac {45R_{\sigma}^4 s^2}{f}+\frac{R_{\sigma}^{12}
s^6}{f^3}+\frac {15R_{\sigma}^8 s^4}{f^2}\right\}-\right.\\ \left.-6s^2\left\{ 15+ \frac{R_{\sigma}^8 s^6}{f^2}+\frac{10R_{\sigma}^4 s^2}{f}\right\}+\right. \\ \left.
+\frac{s^2 c^2( R_{\mu}^2-R_{\sigma}^2)}{f} \left\{ \frac {21R_{\sigma}^8s^4}{f^2}+\frac {R_{\sigma}^{12} s^6}{f^3}+105+\frac {105R_{\sigma}^4s^2}{f} \right\} \right), \\
\end{gathered}
\end{equation}
\begin{equation}\label{AppA:5}
\begin{gathered}
r_{2}=\frac{FR^5R_{\sigma}^2c^2}{48fR_{\mu}^4},\\
\end{gathered}
\end{equation}
\begin{equation}\label{AppA:6}
\begin{gathered}
r_{3}=\frac{FR^{5}R_{\sigma}^2}{48f^2R_{\mu}^2}\left(c^2 \left\{3+\frac{6R_{\sigma}^4 s^2}{f}+\frac{R_{\sigma}^8 s^4}{f^2}\right\}-4s^2 \left\{3+\frac {R_{\sigma}^4 s^2}{f} \right\}+\right. \\ \left. + \frac{s^2 c^2 (R_{\mu}^{2}-R_{\sigma}^2)}{f} \left\{15+\frac{R_{\sigma}^8 s^4}{f^2}+\frac {10R_{\sigma}^4 s^2}{f}\right\} \right),\\
\end{gathered}
\end{equation}
\begin{equation}\label{AppA:7}
\begin{gathered}
r_{4} =\frac{FR^5R_{\sigma}^2}{16fR_{\mu}^4}\left(c^2\left\{1+\frac{R_{\sigma}^4 s^2}{f} \right\}-2s^2+\frac{s^2c^2(R_{\mu}^2-R_{\sigma}^2)}{f}\left\{ 3+\frac {R_{\sigma}^4s^2}{f} \right\}\right).
\end{gathered}
\end{equation}
Here
\begin{equation}
\begin{gathered}
s=\sin\theta, \quad c=\cos\theta, \quad f=R_{\sigma}^2 s^2+R_{\mu}^2 c^2, \\
 F=\frac{J\epsilon\Lambda R}{\sigma\mu\sqrt{2\pi f}}\exp\left(-\frac{R_{\sigma}^2 R_{\mu}^2c^2}{2 f}\right),
\end{gathered}
\end{equation}
where $J$ is an ion flux, $\theta$ is an incidence angle. In equation \eqref{ch2:eq5} we neglected by the terms $-v_0$ and $\gamma \partial h/\partial x$ since these terms do not affect on the characteristics of the surface patterns. Moreover, we can exclude them using the simple transformations \cite{Makeev_2002}.

If we consider the case of normal incidence of ions($\theta=0$), equation \eqref{ch2:eq5} takes the form
\begin{equation}\label{ch2:eq6}
\begin{gathered}
\frac{\partial h}{\partial t}=\mu (\nabla h)^2 + \nu \nabla^2 h - (D+K) \nabla^4 h - r \nabla^6 h.
\end{gathered}
\end{equation}
The coefficients $\mu, \nu, D, r$ can be calculated from relations \eqref{AppA:1}--\eqref{AppA:7}. Let us note that equation  \eqref{ch2:eq6} does not contain dispersion terms since $\xi_{1,2}=\Omega_{1,2}=\zeta_{1-3}=0$.

\section{NUMERICAL STRATEGY AND COMPUTATIONAL RESULTS}

\begin{figure}[!htb]
\center
\includegraphics[width=140 mm]{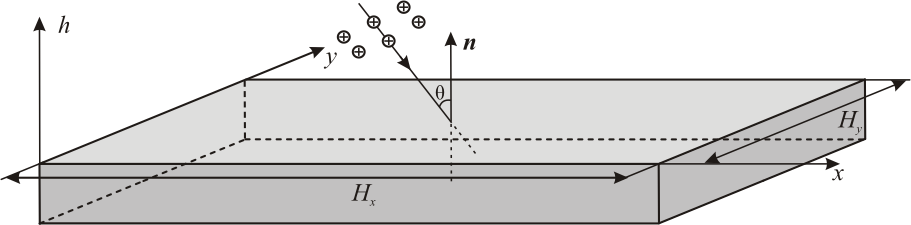}
\caption{{\fontshape{sl}\selectfont  The sputtering processes of the substrate surface}}\label{ff}
\end{figure}

In this section we consider the sputtering processes of semiconductor surface by low energy ion bombardment. The ion flux incident falls on the rectangular substrate surface at the angle $\theta$ from the moment $t=0$ (Fig. \ref{ff}). The energy of incident ions is taken in the interval $\epsilon\subseteq[1, 100]$ keV. The ion flux was taken $J=10^{4} \mu\mbox{A/cm}^2$. The linear dimensions of the substrate are $H_x\times H_y$. Let us suppose that the rectangular area is far enough from the boundary of substrate, allowing us to study the sputtering processes taking into account the periodic boundary conditions.

It is known that the pseudo--spectral time--stepping method is an effective numerical technique that allows us to solve periodic boundary value problems. According to this approach we can rewrite equation \eqref{ch2:eq5} in the following form
\begin{equation}\label{ch3:eq1}
\begin{gathered}
h_t=\emph{\textbf{L}}[h]+\emph{\textbf{N}}[h],
\end{gathered}
\end{equation}
where $\emph{\textbf{L}}, \emph{\textbf{N}}$ are linear and nonlinear operators respectively. Using the Fourier transform algorithm and taking into account periodic boundary conditions we can reduce our problem to the following system of ordinary differential equations
\begin{equation}\label{ch3:eq2}
\textbf{\textit{H}}_t=\mathcal L \textbf{\textit{H}}+\mathcal N[\textbf{\textit{H}}],
\end{equation}
where $\textbf{\textit{H}}, \mathcal L, \mathcal N$ are Fourier transforms of $h, L, N$ respectively. To solve the problem \eqref{ch3:eq2} we use the mixed semi-implicit third order Adams--Bashfort and Crank--Nicholson scheme \cite{Boyd, KRFK_2013, Nepomnyashchy}. Since the Adams--Beshfort algorithm is a multistep method, we cannot use it to calculate the first two layers. So  we used the integrating factor with fourth order Runge--Kutta method \cite{KudRyabSin_2011}to overcome this difficulty. To decrease the magnitude of the error associated with the approximation of nonlinear terms by the Fourier transform we use  \textquotedblleft$2/3$\textquotedblright -- rule. The numerical algorithm was tested on exact solutions obtained in \cite{KR2014, KRFK_2013, KudMig_2007, KudRyab_2011}. Note that equations \eqref{ch2:eq5}, \eqref{ch2:eq6} were studied in dimensionless case. We divided all spatial variables on the average penetration depth $R$ and introduced new variables $h'=h/R, x'=x/R, y'=y/R, t'=t/t_0$ where $t_0=1$ s.

To demonstrate the influence of high order terms on the evolution of the surface height we used three criteria. Firstly, we calculated the evolution of the average surface height function denoted as
\begin{equation}
h_{av}(t)=\frac{1}{H_xH_y}\sum_{x,y}^{H_x,H_y}h(x,y,t)
\end{equation}
Secondly we plotted the evolution of the root mean square deviation of function $h(x,y,t)$ to analyze the surface roughness evolution. We used the following formulae
\begin{equation}
W(t)=\left\langle\sqrt{\frac{1}{H_xH_y}\sum_{x,y}^{H_x,H_y}h^2(x,y,t)-h_{av}^2}\right\rangle,
\end{equation}
where $\langle \cdot \rangle$ is smoothing using the method of least squares. And finally we performed qualitatively comparison of the obtained surface morphologies.

Let us discuss the similarities in the description of the sputtering process using the fourth and sixth order models for normal and oblique incidence of ions.
\begin{figure}[h]
\center
\includegraphics[width=100 mm]{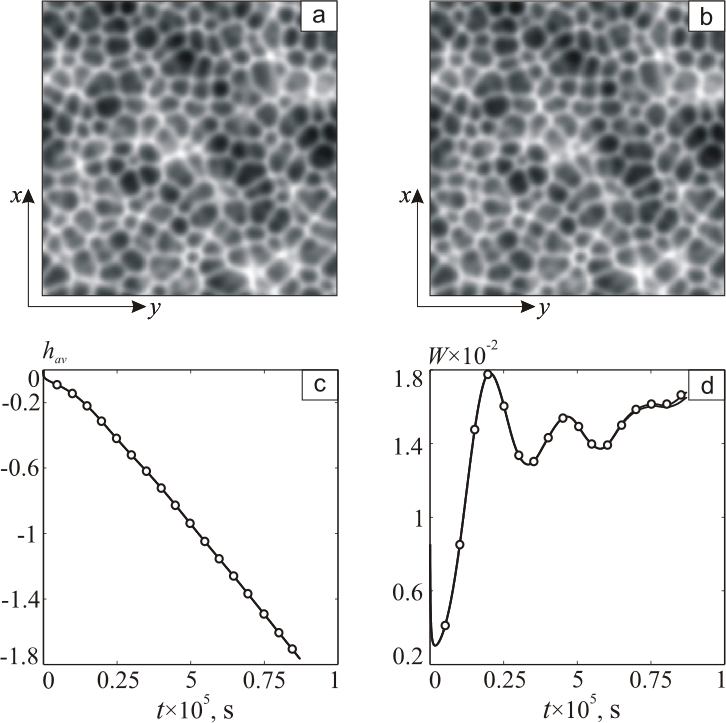}
\caption{{\fontshape{sl}\selectfont  Normal incidence of ions ($\theta=0, R_{\sigma}=2, R_{\mu}=6, K=0.2$). (a) the fourth order model ($t=10^5$); (b) the sixth order model ($t=10^5$); (c) an average height of the surface; (d) surface roughness (root mean square deviation). Doted line is the sixth order model, dashed line is the fourth order model.}}\label{fff}
\end{figure}

\begin{figure}[h]
\center
\includegraphics[width=100 mm]{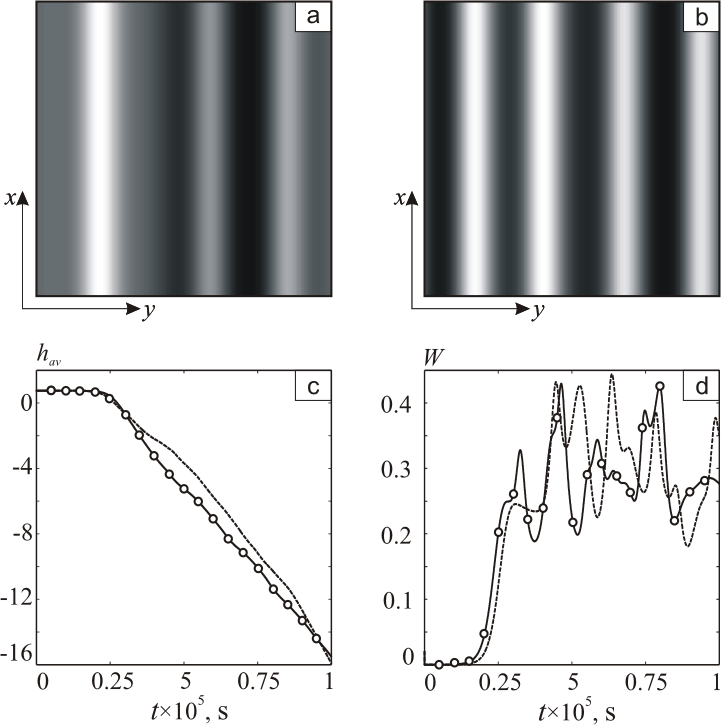}
\caption{{\fontshape{sl}\selectfont Oblique incidence of ions ($\theta=50^{\circ}, R_{\sigma}=2, R_{\mu}=4, K=2.5$). (a) the fourth order model ($t=10^5$); (b) the sixth order model ($t=10^5$); (c) an average height of the surface; (d) surface roughness (root mean square deviation). Doted line is the sixth order model, dashed line is the fourth order model.}}\label{ffff}
\end{figure}

The computational results show that at the earlier stage of surface morphology formation, when $t < \tau$, the value of the average surface height is equal to zero and the surface roughness growth exponentially. The following behavior is observed for both cases of normal and oblique incidence of ions. This stage of the sputtering process can be correctly described by the liner theory. The theoretical estimation of the crossover time $\tau$ is given in \cite{Park1999}. The evolution of $h_{av}(t)$ and $W(t)$ with time is illustrated in Fig. \ref{fff}(c) and (d), \ref{ffff}(c) and (d). From these figures we can see that for both the fourth order and the sixth order models the curves describing the evolution of $h_{av}(t)$ and $W(t)$ practically coincide on this stage. The additional high order terms do not significantly affect on the surface morphology on this stage of sputtering process.

The next stage  of the surface evolution ($t \geqslant \tau$) is characterized by the sharp decrease of the average surface height that corresponds to the beginning of the rebuilding of the surface morphology. From that moment the surface roughness function $W(t)$ begins to oscillate. On this stage the sputtering process can not be properly described by the linear theory. Thus the nonlinear terms become important.

As it turned out, there are some differences in the description of the sputtering process for normal and oblique incidence of ions. In the case of normal incidence of ions the average value of surface height and its roughness coincides on the whole computational domain, see Fig. \ref{fff}(c) and (d). The qualitative profiles of the surface for the fourth and the sixth order model are almost the same. An example of the following morphologies is presented in Fig. \ref{fff}(a) and (b). Thus it is not necessary to take into account the high order terms when we consider the normal incidence case.

However, in the case of oblique incidence of ions the situation changes. Several numerical experiments show that there is no big difference between surface morphologies on the short time scales. However, with time, this small differences become dominant and as we can see from the obtained results, qualitative and quantitative behavior of the numerical solution obtained using the models of the fourth and the sixth order differs, see Fig. \ref{ffff}(a) and (b). The obtained results show that additional high order terms increase the value of the crossover time $\tau$ and change the wave length of the patterns. Also they change the value of the average surface height $h_{av}$. Its absolute value decreases approximately on $10-25 \%$. The surface roughness functions in that case do not coincide that is illustrated in Fig. \ref{ffff}(d). The example of the obtained surface morphology is given in Fig. \ref{ffff}(a) and (b).

\section{CONCLUSION}

In this Letter we have studied the processes of pattern formation on the semiconductor surfaces by low energy ion bombardment. Using the technique proposed in work \cite{Makeev_2002} we have derived a two dimensional sixth order nonlinear partial differential equation that describes the evolution of the surface morphology with time. In fact this equation represents the natural generalization of the Kuramoto--Sivashinsky equation with additional fifth and sixth order terms.

The numerical algorithm based on the pseudo--spectral discretization of the spatial variables and semi implicit-explicit Crank-Nikolson-Adams-Beshfort methods was presented. This algorithm was tested on exact solutions and the results of other authors. Using the proposed numerical strategy we have performed the simulation of the pattern formation processes. We have found that the additional high order terms affect on the final surface morphology when the ion flux incident on the surface at an oblique angle. The value of the average surface height changes approximately on $10-25\%$. However, if we consider the normal incident case, the high order terms can be omitted, since they do not affect to the quantitative and qualitative characteristics of the surface topography.

\section{ACKNOWLEDGMENT}
This work was supported by the RFBR grants 14-01-31078, 14-01-00493 and grant for Scientific Schools 2296.2014.1.

\end{document}